\documentclass[12pt,a4paper,english]{article}
\usepackage[T1]{fontenc}
\usepackage[latin1]{inputenc}
\usepackage{babel}
\usepackage{graphics}

\makeatletter

\providecommand{\LyX}{L\kern-.1667em\lower.25em\hbox{Y}\kern-.125emX\@}

\usepackage{epsfig}
\date{\today}
 \large\normalsize
\newcommand{\bmat}{\left(\begin{array}}
\newcommand{\emat}{\end{array}\right)}

\def\lsim{\raise0.3ex\hbox{$\;<$\kern-0.75em\raise-1.1ex\hbox{$\sim\;$}}}
\def\gsim{\raise0.3ex\hbox{$\;>$\kern-0.75em\raise-1.1ex\hbox{$\sim\;$}}}

\def\LEP2{{LEPII}}

\makeatother
\begin{document}
\begin{flushright}
\small
\end{flushright}
\vspace{.3cm}

\begin{center}
{\bf{\Large
CP violation versus flavour in supersymmetric theories}}\\
\vspace* {1cm}
{\large
S. Abel\( ^{1} \), G. C. Branco\( ^{2} \) and S. Khalil\( ^{1,3} \)}\\
\vspace* {5mm}
\( ^{1} \) \emph{IPPP, Centre for Particle Theory, Durham
University, DH1 3LE, Durham,~~U.~K.}\\
 \( ^{2} \) \emph{Centro de F\'{\i}sica das Interac\c{c}\~{o}es
Fundamentais (CFIF), Departamento de F\'{\i}sica, Instituto Superior
T\'{e}cnico, Av. Rovisco Pais, 1049-001 Lisboa, Portugal}\\
 \( ^{3} \) \emph{Ain Shams University, Faculty of Science, Cairo,
11566, Egypt.}

\vspace*{10mm}
{\bf \large Abstract}
\end{center}

We show that the quark flavour structure and CP violating phenomena are strongly correlated 
in supersymmetric theories. For a {\em generic} pattern of supersymmetry breaking 
the two broad categories of Yukawa couplings, democratic and hierarchical textures, have entirely 
different phenomenological implications. With hierarchical Yukawas, the rephasing invariant
phase, $\mathrm{arg}(V_{us} V_{cb} V_{cb}^* V_{cs}^*)$, in the CKM mixing 
matrix has to be of order unity, while the SUSY CP violating phases are 
severely constrained by electric dipole moments, giving rise to the so-called SUSY
CP problem. With democratic Yukawas, all experimental CP results 
can be accommodated with small values for the CKM and SUSY CP violating phases ({\it i.e.}, CP can 
be considered as an approximate symmetry {\em at the high energy scale}). We also show that within 
this scenario, an entirely  real CKM matrix in supersymmetric models is still allowed by the present 
experimental results.

\newpage
%
%

\section{{\large \bf Introduction}}

In the Standard Model (SM), CP violation and flavour transition arise
from complex Yukawa couplings which lead to a physical CP violating phase in 
the Cabibbo--Kobayashi--Maskawa (CKM) mixing matrix.
Within the SM, the rephasing invariant phase $\delta_{CKM} \equiv \mathrm{arg} ( V_{us}
V_{cb} V^*_{ub} V_{cs})$ has to be of order unity in order to account for the observed 
CP violation in the kaon sector. Recent experimental
results on CP violation in B decays \cite{babar} are consistent with
the SM and also with the Constrained MSSM with flavour universality,
which requires a similarly large value of \( \delta _{CKM} \). It
is tempting therefore to conclude that the possibility of a real CKM matrix
is now excluded and that CP violation in the Yukawas has been established
and is dominant \cite{nir}. In this paper we demonstrate
that this conclusion is premature. Although the SM and Constrained MSSM
provide consistent predictions of the experimentally determined CP
violation, there remain reasonable supersymmetric models with \( \delta _{CKM}=0 \)
that are also consistent with experiment.

The models in question must necessarily
have large flavour non--universality and yet satisfy constraints on
for example electric dipole moments (EDMs). We will see that reasonable predictions
can still be obtained in models with Yukawas that are close to democratic
and whose (flavour non--universal) SUSY breaking patterns can be motivated
by effective D-brane models. More generally, there is a correlation between 
the flavour structure of the Yukawas and CP violating phenomena. Those models that 
have more democratic Yukawas can have smaller CKM phases, thereby mitigating 
the so called supersymmetric CP problem by reducing the contribution to EDMs.

Democratic textures are one of the two broad categories of Yukawa couplings.
The first category has of course the familiar hierarchical structure \cite{kobayashi}:
\( Y_{33}>Y_{ij},Y_{22}>Y_{mn} \) where \( Y_{ij} \) is any entry
except \( Y_{33} \), and \( Y_{mn} \) denotes the \( (1,1) \),
\( (1,2) \) and \( (2,1) \) entries. This type of Yukawa texture
can be considered as a perturbation around diagonal fermion mass matrices.
The second category includes the Yukawa couplings with nearly democratic
textures which are perturbations around the matrix \begin{equation}
\Delta =\left( \begin{array}{ccc}
1 & 1 & 1\\
1 & 1 & 1\\
1 & 1 & 1
\end{array}\right) .\end{equation}
Although these two classes look quite different and can have very
different motivations, they lead to exactly the same physics within
the framework of the SM. Both can give the correct fermion masses
with the measured values of the CKM mixing matrix. Indeed, within the 
framework of the SM, one can go from the democratic texture to the 
hierarchical one, through a weak basis (WB) transformation which keeps
the gauge current flavour diagonal. Within the SM, the physics does not change 
when one makes the above WB transformation. However this is no longer the case 
when one considers extensions of the SM and in particular supersymmetric ones. 

As a result, we shall see in this paper that, \textit{for a given pattern} of supersymmetry
breaking in supersymmetric extensions of the SM, these two types of
Yukawa have entirely different phenomenological implications. We will
demonstrate that a real CKM matrix is still possible in democratic
models with flavour non-universal A-terms. (Rotating to a
basis in which the Yukawas are hierarchical would yield A-terms with
a very peculiar hierarchical structure.) In models of this type, the
SUSY contributions, as we will show in detail below, can be dominant
and can saturate all the CP experimental results even with vanishing
\( \delta _{CKM} \). (This is a generalization of the conclusions
of ref.\cite{branco} which considered pure phase matrices and found a small
amount of CP violation from the usual KM mechanism and significant
SUSY contributions.)

If instead we take hierarchical Yukawas and keep the A-terms to be
of the same order, the usual KM mechanism must give the dominant contribution
to all CP violating measurements, since it is impossible to account
for the observed CP violation in \( K \) and \( B \) systems by
the supersymmetric contributions alone and without exceeding the limits
from electric dipole moments (EDMs)~\cite{Abel:2001vy}. 
I models of this type we inevitably require \( \delta _{CKM} \) of order unity. 
We should comment that in Ref.\cite{kane}, a supersymmetric model with a real CKM and 
large flavour--independent SUSY soft phases was considered. It was shown that it is possible, 
with unconventional flavour structure of squark mass matrices and order one SUSY 
soft phases, that SUSY contributions can account for the experimental
results of CP violation in the kaon system and $a_{J/\psi}$.
The EDM problem of this model was assumed to be solved by cancellation between 
different contributions. However, as shown in Ref.\cite{Abel:2001vy}, 
this cancellation can not suppress simultaneously the EDM of the electron, neutron and mercury.

Unsuccessful searches for EDMs are without doubt the most constraining
factor for SUSY models that seek to provide alternative explanations
for the CP data, and so we begin in the following section, by reviewing
the various ways in which EDMs might be suppressed. As well as the
conventional Constrained MSSM with flavour universality, there are
a number of other far less restrictive ideas available, such as hermitian
flavour structure, and factorizable A-terms. In particular D brane
models provide a realization of the latter, and can therefore naturally
lead to suppression of EDMs. In the sections that follow, we then
consider the purely supersymmetric production of \( \varepsilon  \)
and \( \varepsilon '/\varepsilon  \) within these frameworks, and
go on to show how the models outlined above can also generate sufficient
CP violation in B mixing.

\section{{\large \bf Suppressing EDMs}}

EDMs are a serious problem for supersymmetric models because of the
copious sources of CP violating phases. A generic SUSY model predicts
values of the neutron, electron and mercury EDMs that are many orders
of magnitude larger than the current bounds \cite{bound, eedm, mercury},\begin{eqnarray}
d_{n} & < & 6.3\times 10^{-26}\mathrm{ecm}(90\%CL),\nonumber \\
d_{e} & < & 4.3\times 10^{-27}\mathrm{ecm},\nonumber \\
d_{Hg} & < & 2.1\times 10^{-28}\mathrm{ecm}.
\end{eqnarray}
 With the expected improvements in experimental precision, the EDM
is likely to be one of the most important constraints on physics beyond
the Standard Model for some time to come. (Indeed one of the reasons
EDMs are so important is that sufficient baryogenesis requires CP
violation beyond that in the Standard Model, and EDMs therefore restrict
the possibilities.) 

The complete expressions valid for any basis were presented in Ref.~\cite{Abel:2001vy}
and we shall use these expressions throughout our analysis. The contributions
can also be understood in terms of leading order mass insertion diagrams,
and this provides a useful model independent constraint on the imaginary
part of the mass insertions. The mercury EDM is typically most constraining
and one finds\begin{eqnarray*}
Im(\delta _{11}^{d,u})_{LR} & < & 7.10^{-8}\\
Im(\delta _{22}^{d})_{LR} & < & 6.10^{-6}
\end{eqnarray*}
 for squark masses of order \( 500 \)GeV.

Any viable model must therefore have a pattern of flavour and/or supersymmetry
breaking that leads to EDM suppression. There are several candidates
patterns which can do this, of which the Constrained MSSM with small
SUSY phases is merely the most restrictive example. Our first task,
which is the subject of this section, is therefore to consider the
general properties of EDMs in SUSY and to identify the patterns that
can naturally lead to their suppression. This will enable us in the
following section to consider those examples that have real CKM matrices.
An additional task, which we only partially address here, is to find
underlying mechanisms which might be responsible for generating these
flavour and supersymmetry breaking patterns.

Let us begin by writing the high energy scale soft breaking potential as
\begin{eqnarray}
V_{SB} & = & m_{\alpha \beta }^{2}\phi ^{*}_{\alpha }\phi _{\beta }-
(B\mu H_{1}H_{2}+h.c.)+(A_{l_{ij}}Y_{ij}^{l}\; H_{1}\tilde{l}_{Li}
\tilde{e}_{Rj}^{*}+A_{d_{ij}}Y_{ij}^{d}\; H_{1}\tilde{q}_{Li}\tilde{d}_{Rj}^{*}\nonumber \\
 & - & A_{u_{ij}}Y_{ij}^{u}\; H_{2}\tilde{q}_{Li}\tilde{u}_{Rj}^{*}+
h.c.)+{1\over 2}(m_{3}\overline{\tilde{g}}\tilde{g}+m_{2}\overline{\tilde{W}^{a}}
\tilde{W}^{a}+m_{1}\overline{\tilde{B}}\tilde{B})\; ,
\end{eqnarray}
 where \( \phi _{\alpha } \) denotes all the scalars of the theory.
In the complete Lagrangian there are now 43 new phases in addition
to the phase of the CKM matrix. In all cases the phase of \( \mu  \)
is dangerous for EDMs (we do not consider cancellation of phases,
although we consider this to be extremely unlikely for the reasons
set out in Ref.~\cite{Abel:2001vy}), so it is reasonable to assume that it is 
automatically set to zero by whatever mechanism is responsible for the creation
of the \( \mu  \) term.

The A-terms are more problematic because of their flavour changing
nature, and in fact both the CP phases \textit{and} flavour structure
need to be specified to ensure suppression of EDMs. Indeed even if
the A-terms are real in the SUGRA basis in which they are calculated,
they may still induce large EDMs if the Yukawa couplings are not almost
diagonal in that basis~\cite{Abel:2001cv}. Generally therefore, the conventional SUSY
flavour and CP problems are inextricably linked with each other and
with the Yukawas, and EDMs provide a restrictive constraint on \textit{the
entire flavour structure} of generic fundamental models, the so-called
string CP problem~\cite{Abel:2001cv}.

There are a number of possible flavour and/or SUSY breaking patterns
which may explain the absence of additional contributions to EDMs:

\begin{itemize}
\item Flavour universality and small phases in SUSY breaking: This is the
most severe assumption but works independently of the Yukawa structure.
It is allowed by for example dilaton dominated breaking in supergravity
models. However the latter possibility is even more restrictive and
appears to be ruled out by both experimental and cosmological considerations.
In addition dilaton domination does not occur very naturally when
one considers the dynamical stabilization of dilaton and moduli fields. 

\item Approximate CP: If we relax the assumption of flavour universality
then generally we will get large EDMs. We repeat that this is the
case even if the A-terms are real in the supergravity basis where
they are calculated, because EDMs are calculated in the mass basis
and the rotation to this basis will generally introduce phases of
the order of \( \delta _{CKM} \) . Thus in models with a large \( \delta _{CKM} \)
the dipole moments are a constraint on the flavour structure of the
A-terms. The approximate CP idea seeks to avoid this problem by making
all phases including \( \delta _{CKM} \) small (of order \( 10^{-2} \)
or less). This is enough to suppress EDMs below the current experimental
bound, but the experimentally observed CP violation in the K and B
systems must then be made up chiefly from supersymmetric contributions. 

\item Hermitian flavour structure: Large phases are allowed if they do not
appear in the flavour diagonal SUSY breaking in the basis where we
are calculating the EDMs. It is unreasonable to expect this to be
the case unless the supersymmetry breaking is flavour universal (in
which case any change of basis commutes with the A-terms) or the flavour
structure is hermitian, with CP violation being associated with flavour
off-diagonal phases~\cite{Abel:2000hn}. If both the Yukawas and A-terms are hermitian,
then they remain so in any basis and the contribution to EDMs is extremely
small (i.e. it is induced only by renormalization group effects).
A hermitian mass matrix is difficult to achieve however, but it can
arise in models where the flavour structure is associated with the
VEVs of higgs fields in the adjoint representation of an SU(3) flavour
group~\cite{Khalil:2002jq}. An added benefit of this association of CP violation 
with flavour structure is that it explains why there is no phase on the \( \mu  \)
term. 

\item Factorizable A-terms. A rather less dramatic assumption is that the
A-terms are factorizable; that is they can be written\begin{equation}
A_{ij}Y_{ij}=(\underline{a}.Y)_{ij}\end{equation}
 where \( \underline{a} \) is some matrix. It is not hard to see
why this leads to a suppression of EDMs if we examine a typical contribution
to the down-quark EDM from the gluino diagram. At leading order
this contribution is proportional to the A-term mass insertion above. To
calculate it we rotate to the mass basis, where the mass insertion
becomes\begin{equation}
Im\left( (S_{L}^{\dagger }.\underline{a}.S_{L})_{11}\frac{m_{d}}{v_{1}}\right) .\end{equation}
 where \( S_{L} \) is the diagonalization matrix of the left handed
quarks. The situation is similar to that in the Constrained MSSM with
non-zero phases. There is a partial suppression due to the up-quark
insertion, but satisfying EDM bounds requires a further suppression
which can be achieved by setting a single phase to be smaller than
\( 10^{-2} \). In the Constrained MSSM the phase in question is that
of the universal A-term, whereas in the factorizable case the phase
is that of \( a_{11} \). If there is a large amount of mixing as
in the democratic case there will also be constraints on the phases
of the other elements of \( \underline{a} \), but depending on \( S_{L} \)
this is typically less severe (\( 10^{-1} \) for example). For chargino
diagrams there is an additional mixing due to the CKM matrix, but
an additional suppression due to the weak interaction vertices. Factorizable
A-terms arise rather naturally in D-brane models as explained in 
Ref.~\cite{Khalil:2000ci}, where typically the A-terms will have degenerate rows 
\begin{equation}
A_{ij}=\left( \begin{array}{ccc}
a & a & a\\
b & b & b\\
c & c & c
\end{array}\right) \, \, \rightarrow \, \, \underline{a}=\left( \begin{array}{ccc}
a &  & \\
 & b & \\
 &  & c
\end{array}\right) .\end{equation}
When the Yukawas are hierarchical, EDM suppression only requires 
that the phase \( \theta _{a} <10^{-2} \), and leaves \( \theta_b \) 
and \( \theta_c \) essentially unconstrained.
In the democratic case the mixing to the third generation from \(S_L\) 
is significant, and one typically requires \( \theta_a, \, \theta_c < 10^{-2} \) 
and \( \theta_b < 10^{-1}\).
\end{itemize}
It is clear that the last two patterns may still allow real CKM matrices.
The fact that the supersymmetric contributions to CP violating processes
must be dominant implies that there should be large phases somewhere
in the model which implies that the hermitian and factorizable 
patterns may still be viable with real CKMs. Rather surprisingly 
however, when we take the approximate CP limit of small phases, 
we can {\em still} find models that are viable with 
democratic Yukawas. As we shall see, this is because small phases at the 
GUT scale can be transformed into large ones at the weak scale by 
renormalization group effects when the mixing is large. 
   In order to get significant contributions
to the CP asymmetry observed in B decays, the supersymmetric 13 mixing
should be large as well. All of these conditions can be met by models
that have real and almost democratic Yukawas, and either hermitian
or factorizable A-terms, even in the approximate CP limit, as we shall 
now see. 

We will begin in this section by considering EDMs and confirming the 
general behaviour described above, and then 
in the following sections go on to consider the other CP violating parameters.
For our numerical estimates, it will be useful to have the following specific
example of Yukawas in mind; 
\begin{eqnarray}
Y_{u} & = & \frac{1}{v\sin \beta }\mathrm{diag}\left( m_{u},m_{c},m_{t}\right) ,\nonumber \\
Y_{d} & = & \frac{1}{v\cos \beta }K^{T}.\mathrm{diag}\left( m_{u},m_{c},m_{t}\right) .K^{*},
\label{Yherar}
\end{eqnarray}
 where \( K \) is the CKM matrix. This texture is a standard example
of hierarchical Yukawas. The CKM matrix \( K \) is formed from the
unitary transformations that diagonalize the mass matrices in the
up and down sectors which in this case are given by \( S^{u}_{L}=S^{u}_{R}=I \)
and \( S^{d}_{L}=S^{d}_{R}=K \). This shows that the hierarchy inherent
in the Yukawa textures reveals itself in these rotation, which as
we will discuss below has important consequences in the SUSY results.
Conversion of the hierarchal Yukawa matrices to democratic ones can
be brought about the unitary transformation \begin{equation}
Y_{\mathrm{dem}}=UY_{\mathrm{her}}U^{+}
\end{equation}
 where \( U \) is given by \begin{equation}
U=\left( \begin{array}{ccc}
\frac{1}{\sqrt{6}} & \frac{1}{\sqrt{6}} & -\frac{2}{\sqrt{6}}\\
-\frac{1}{\sqrt{2}} & \frac{1}{\sqrt{2}} & 0\\
\frac{1}{\sqrt{3}} & \frac{1}{\sqrt{3}} & \frac{1}{\sqrt{3}}
\end{array}\right) .
\end{equation}
 In this case, we find that \( Y^{u}_{\mathrm{dem}} \) and \( Y^{d}_{\mathrm{dem}} \)
are given by \begin{eqnarray}
Y^{u}_{\mathrm{dem}} & = & \frac{\lambda _{u}}{3}\left( \begin{array}{ccc}
1.013 & 0.987 & 0.999\\
0.987 & 1.013 & 0.999\\
0.999 & 0.999 & 1
\end{array}\right) ,
\label{yuku}\\
Y^{d}_{\mathrm{dem}} & = & \frac{\lambda _{d}}{3}\left( \begin{array}{ccc}
0.987 & 0.905 & 0.968\\
0.903 & 1.212 & 1.008\\
0.967 & 1.008 & 1
\end{array}\right) ,
\label{yukd}
\end{eqnarray}
 where \( \lambda _{u}=m_{t}/v\sin \beta  \) and \( \lambda _{d}=m_{b}/v\cos \beta  \).
Numerically, the Yukawa couplings are now diagonalized by the transformations
\begin{eqnarray}
S^{u}_{L} & \simeq  & S^{u}_{R}\simeq \left( \begin{array}{ccc}
-0.408 & -0.408 & 0.816\\
0.707 & -0.707 & 0\\
-0.577 & -0.577 & -0.577
\end{array}\right) ,\label{sud}\\
S^{d}_{L} & \simeq  & S^{d}_{R}\simeq \left( \begin{array}{ccc}
-0.557 & -0.246 & 0.793\\
0.622 & -0.757 & 0\\
-0.551 & -0.606 & -0.574
\end{array}\right) .
\end{eqnarray}
Evidently in this case the matrices \( S^{u,d}_{L,R} \) have large
mixing, and in particular the mixing between the first and the third
generation is much larger than it is in the case of hierarchical Yukawas
where \( (S^{u}_{L,R})_{13}=0 \) and \( (S^{d}_{L,R})_{13}=K_{13}\sim 10^{-3} \).
In the SM only \( S^{u}_{L}.S^{d^{\dagger }}_{L}\equiv K \) is physically
meaningful, but in SUSY models, in particular with non--universal
soft breaking terms, these matrices play a significant role as we
will discuss below.

We now need to assume one of the patterns of \( A \) terms specified
above, that avoid large EDMs whilst simultaneously enhancing the SUSY
contributions to the other CP observables; {\em i.e.} we will assume either
that the flavor structures are completely hermitian (\( Y^{q}=Y^{q^{\dagger }} \)
and \( A^{q}=A^{q^{\dagger }} \)) or that the A-terms are matrix
factorizable (\( \hat{A}=A.Y \) or \( Y.A \)). In the former case,
flavour blind quantities such as the \( \mu  \)--term and the gaugino masses
are real while in the latter this has to be assumed. Also since
in eqs.(\ref{Yherar},\ref{yuku},
\ref{yukd}) we consider symmetric and real Yukawas the
hermiticity assumption requires simply that our \( A \)-terms are
hermitian. Note that these choices affect only the EDMs and do not
greatly affect our later conclusion of dominant supersymmetric contribution
to the other CP violating parameters such as $\varepsilon $ and $\varepsilon' $, 
provided that the Yukawas are democratic. Indeed we can obtain similar results 
for any other choice of non--universal \( A \)--terms (but of course with general 
overproduction of EDMs). 

Now consider the following hermitian \( A \)--terms;\begin{equation}
\label{hermitA}
A_{d}=A_{u}=\left( \begin{array}{ccc}
A_{11} & A_{12}e^{i\varphi _{12}} & A_{13}e^{i\varphi _{13}}\\
A_{12}e^{-i\varphi _{12}} & A_{22} & A_{23}e^{i\varphi _{23}}\\
A_{13}e^{-i\varphi _{13}} & A_{23}e^{-i\varphi _{23}} & A_{33}
\end{array}\right) \; .
\label{Aterms}
\end{equation}

\begin{figure}
{\centering \begin{tabular}{cc}
\resizebox*{0.45\textwidth}{0.2\textheight}{\includegraphics{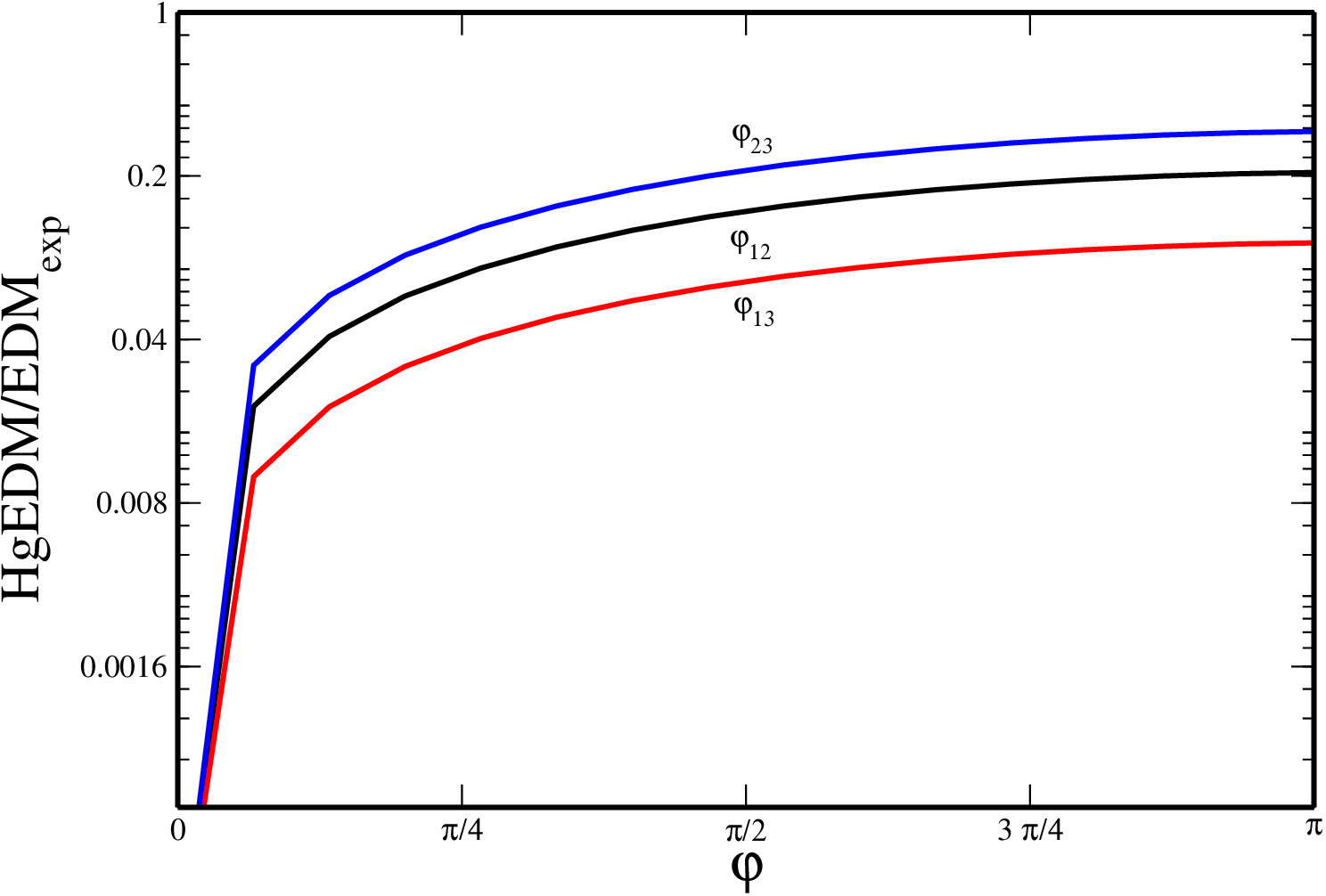}} &
\resizebox*{0.45\textwidth}{0.2\textheight}{\includegraphics{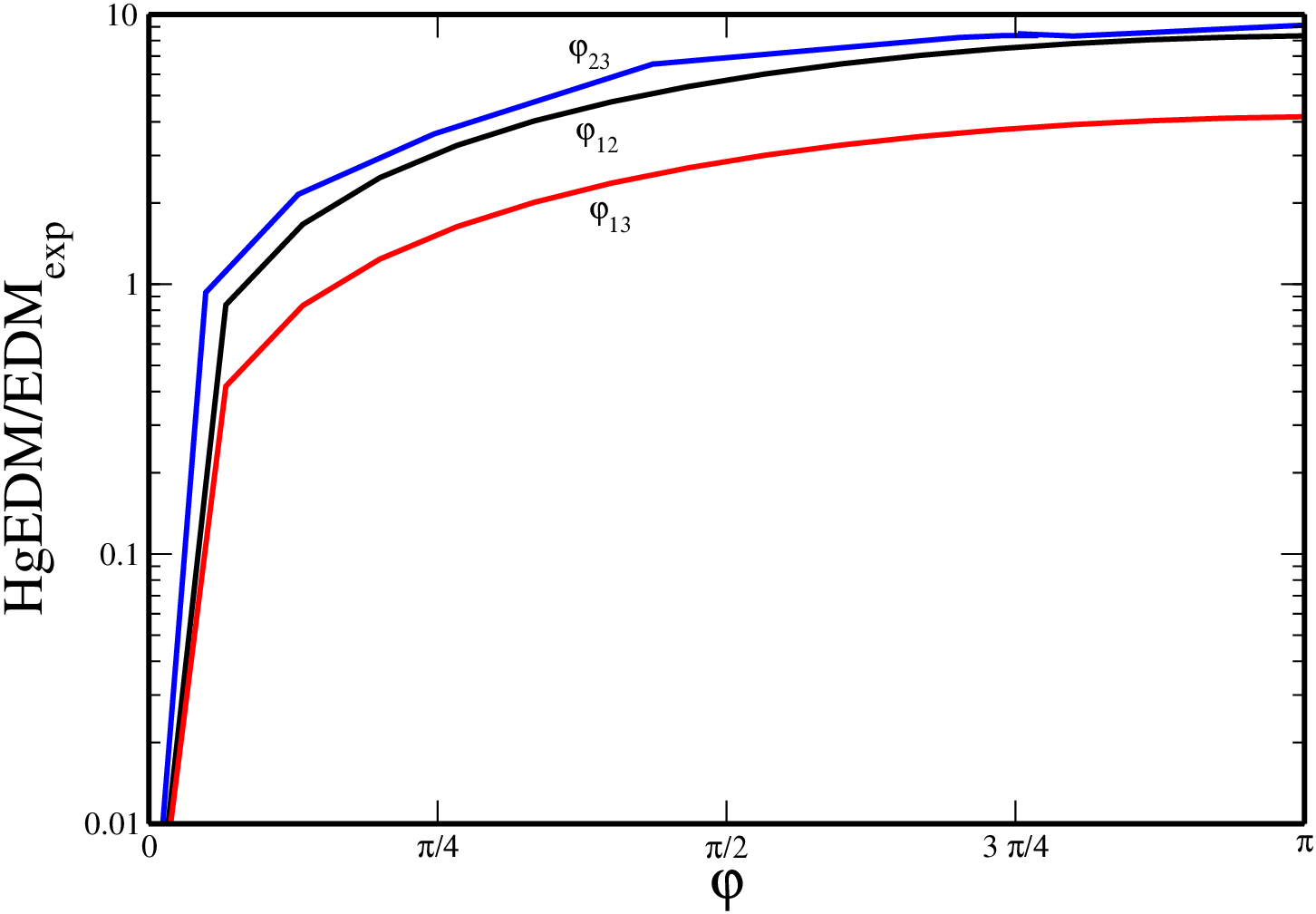}} \\
\end{tabular}\par}
\caption{\label{fig1}The mercury EDM versus the off-diagonal phases of the
hermitian A-terms for hierarchical (left) and democratic (right) Yukawas.}
\end{figure}

In Figure 1, we display the mercury EDMs as a function of the flavour
off--diagonal phases of the \( A \)--terms in Eq.(\ref{hermitA})
with the hierarchal and nearly democratic Yukawa examples given in eqs.(\ref{Yherar},\ref{yuku},
\ref{yukd}).
We assume \( \tan \beta =5 \), \( m_{0}=m_{1/2}=250 \) GeV, \( A_{ii}=m_{0} \),
\( A_{12}=-2m_{0} \), \( A_{13}=-m_{0} \) and \( A_{23}=2m_{0} \)
As noted in ref.\cite{Khalil:2002jq}, in the case of hierarchal Yukawas the EDM
limits do not impose significant constraints on the phases of the
hermitian \( A \)--terms since the entire contribution is from renormalization
group running. However, with nearly democratic Yukawas, because
of the significant mixing between the different generations, we observe
that the strongest constraint coming from the mercury EDM requires
that the off--diagonal phases be less than \( \pi /10 \).

As mentioned earlier, because of the large mixing, the bounds
on the phases of the factorizable \( A \)--terms with democratic
Yukawas are more stringent than that in the case of hierarchal Yukawas. 
As found in ref.\cite{branco}, with universal strength Yukawas compatibility with EDMs
constrains these phases to be of order \( 10^{-2}-10^{-1} \), whereas
with hierarchal Yukawas some phases can be of order one.

\section{{\large \bf\protect\( \varepsilon_K \protect \) and \protect\( \varepsilon '/\varepsilon \protect \)
with a real CKM matrix}}

Generally one expects a considerable enhancement of CP violating processes 
in supersymmetric extensions of the SM by both new SUSY CP violating 
phases and also by new flavor structures. 
However in the most constrained case of SUSY models with minimal flavor
violation (as in mSUGRA, where universality of the soft SUSY
breaking terms is assumed and the only source of flavour structure
is the Yukawa matrices),
the two physical SUSY phases are constrained by the EDMs to be 
\( \mathcal{O}(10^{-2}) \). If in addition \( \delta _{CKM}=0 \), the SUSY 
contributions cannot account for CP violating measurements
such as \( \varepsilon _{K} \), \( \varepsilon' /\varepsilon  \) and
the CP asymmetry of the \( B^{0} \) mesons. This is true
for both types of Yukawa since, as mentioned above, due the universality of 
the soft breaking terms the matrices \( S^{u,d}_{L,R} \) have no effect. 
For instance the LR part of the squark mass matrix in the super-CKM basis is 
given by \( S_{L}^{q}\hat{A}^{q}S^{q\dagger }_{R} \), where 
\( \hat{A}^{q}_{ij}=Y^{q}_{ij}A^{q}_{ij} \).
Thus for universal trilinear couplings we obtain 
\( AS_{L}^{q}Y^{q}S^{q\dagger }_{R}=AY^{d}_{\mathrm{diag}} \).
So, as in the SM, the matrices \( S^{u,d}_{L,R} \) do not play any role.

This fact has motivated a growing interest in SUSY models with non-universal 
soft breaking terms~\cite{Abel:1996eb}. It has been proven that the
trilinear couplings play an important role and that new flavour structure
in the \( A \) terms can saturate the experimental
measurement of \( \varepsilon _{K} \) and \( \varepsilon '/\varepsilon  \). 

It is remarkable that, although EDMs are more constraining for 
the SUSY phases in the case of the democratic Yukawa couplings, 
the SUSY contribution
to the CP observables in the \( K \) system are significant
and much larger than the SUSY contributions with large SUSY phases
and very small mixing between generations \cite{Khalil:2002jq}. In the democratic class
of models the SUSY contribution can easily saturate the experimental
values of \( \varepsilon _{K} \) and \( \varepsilon' /\varepsilon  \).
Indeed, it has been shown in ref.\cite{Khalil:2002jq} that the gluino mediated boxes with
\( LL \) mass insertions give the leading contributions to \( \varepsilon _{K} \)
and that \( \varepsilon' /\varepsilon  \) is dominated by the chargino
loops with \( LL \) mass insertions.\\

\begin{figure}[t]
\begin{center}
\epsfig{figure=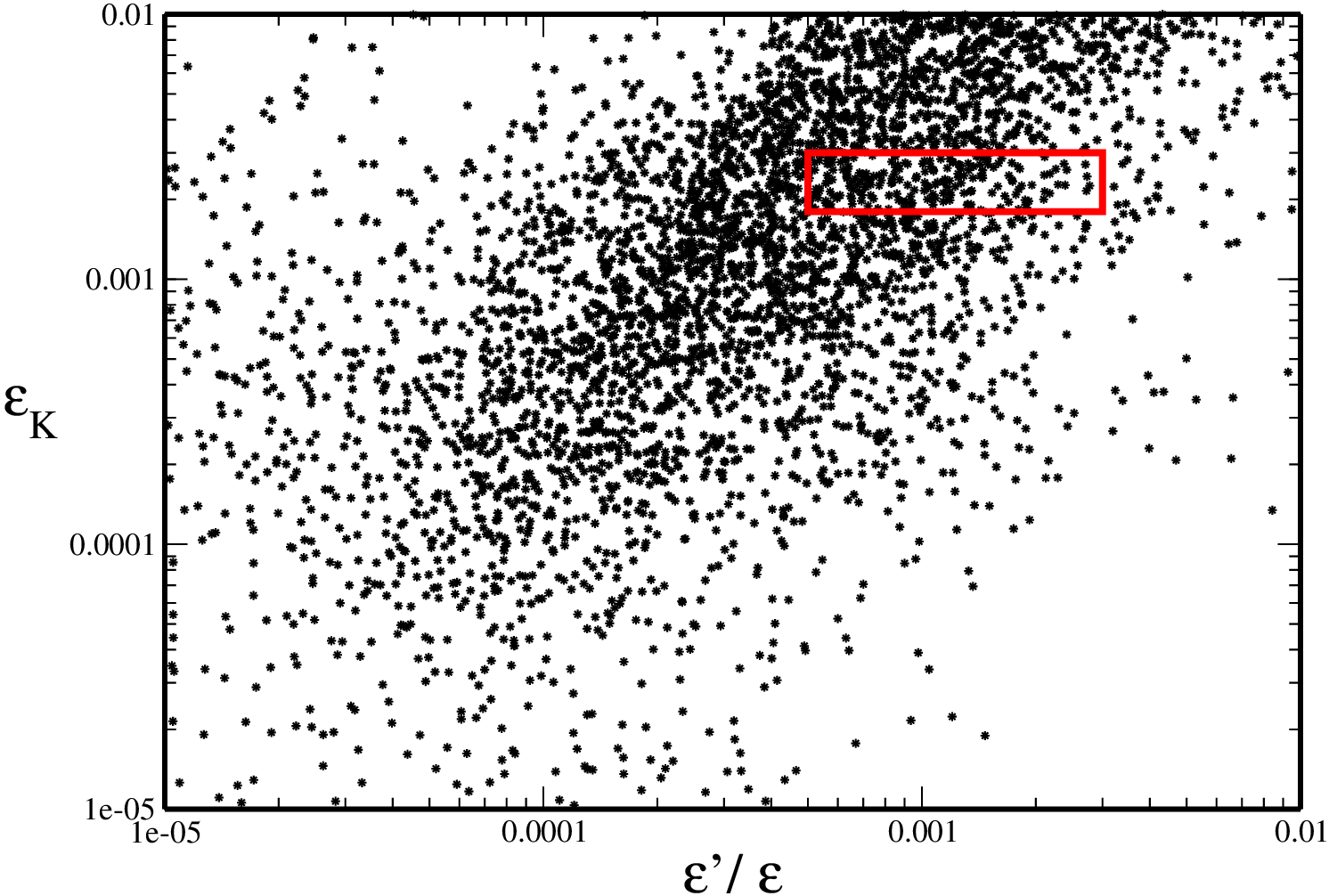,height=6.25cm,width=10.5cm,angle=0}
\end{center}
\caption{Correlation between $\varepsilon_K$ and $\varepsilon'/\varepsilon$ for 
$\tan \beta=5$, $m_0=m_{1/2}=250$ GeV, $A_{ij}\in [-3,3]~m_0$ and 
$\phi_{A_{ij}} \lsim 0.1$. The box represents the SM and the constrained MSSM
results.}
\label{fig2}
\end{figure}

In figure \ref{fig2} we present a scatter plot for the correlation between
the CP violating parameters $\varepsilon_K$ and $\varepsilon'/\varepsilon$
in SUSY model with the democratic Yakawa couplings as in eqs.(\ref{yuku},\ref{yukd})
and Hermitian $A$-terms as in eq.(\ref{Aterms}) for $\tan \beta=5$, 
$m_0=m_{1/2}=250$ GeV, $A_{ij}\in [-3,3]~m_0$ and $\phi_{A_{ij}} \lsim 0.1$.
In this figure we also show the prediction of the constrained MSSM for this
correlation, which as we emphasized is essentially the SM result. 

These results already tell us that a sizable SUSY contribution to 
\(\varepsilon \) and \(\varepsilon'\)
does not require large phases (either in the CKM or in the soft
SUSY terms) but it does require large flavour mixing in the Yukawas together with
flavour structure in the soft terms (such as non--degenerate 
\( A \)-terms). In this case, the CP
violation can act as a probe of the flavour structure of the supersymmetric
theories. This scenario for CP violation, where
the CP violation in the \( K \) and \( B \) systems is fully
supersymmetric, is complementary to the usual one in 
all of the CP phenomena originate from the SM and there is no new flavor structure
beyond the Yukawa. The latter minimal flavour assumption remains valid of course, 
however it does not differ significantly from the SM.


\section{{\large \bf Large \protect\( a_{J/\psi K_S} \protect \) with 
real CKM and the unitary triangle}}

It is tempting to assume that the recently measured large 
CP asymmetries in \( B_{d} \) and \( \bar{B}_{d} \)
meson decay to \( J/\psi K_{S} \) \( (a_{J/\psi K_{S}}) \) observed
by BaBar and Belle \cite{babar} indicate that \( \delta _{CKM} \)
is of order one. If nature is supersymmetric, then this would lead 
to the very important conclusion that
CP must already be violated by the supersymmetric part of the theory 
({\em i.e.} the superpotential). However
in ref.\cite{emidio} it has been emphasized that the supersymmetric chargino 
contribution to \( B_{d}-\bar{B}_{d} \) mixing can accommodate
this large value of \( a_{J/\psi K_{S}} \) and that the SM contribution
can be relatively small.

In the presence of SUSY contributions the CP asymmetry parameter
is given by \begin{equation}
a_{J/\psi K_{S}}=\sin 2\beta ^{\mathrm{eff}}=\sin (2\beta ^{\mathrm{SM}}+
2\theta _{d}),
\end{equation}
where \begin{equation}
\beta ^{\mathrm{SM}}=\mathrm{arg}\left( -\frac{K_{cd}K_{cb}^{*}}{K_{td}K_{tb}^{*}}\right) \end{equation}
 which equals zero in our case with real \( K \) and \begin{equation}
2\theta _{d}=\mathrm{arg}\left( 1+\frac{M_{12}^{\mathrm{SUSY}}(B_{d})}{M_{12}^{\mathrm{SM}}(B_{d})}\right) .\end{equation}
 The important SUSY contribution to the \( B_{d}-\bar{B}_{d} \) is
given by \begin{equation}
M_{12}^{\mathrm{SUSY}}(B_{d})=M_{12}^{\tilde{g}}(B_{d})+M_{12}^{\tilde{\chi }^{\pm }}(B_{d}).\end{equation}

In order to saturate the experimental values of \( a_{J/\psi K_{S}} \)
through the gluino exchange one should have, for squark mass of order 400 GeV
and gluino mass of order 200 GeV, the following mass insertions~\cite{emidio} 
\begin{eqnarray}
&&\sqrt{\vert \mathrm{Im}[(\delta^d_{LL})_{31}^2]\vert}= 5.2\times 10^{-2},
~~~~~~~~~~~~~~~\sqrt{\vert \mathrm{Im}[(\delta^d_{RL})_{31}^2]\vert}= 2.5\times 10^{-2},\\
&&\sqrt{\vert \mathrm{Im}[(\delta^d_{LL})_{31} (\delta^d_{RR})_{31}]\vert}=
9.6\times 10^{-3},
~~~~~~\sqrt{\vert \mathrm{Im}[(\delta^d_{LR})_{31} (\delta^d_{RL})_{31}]\vert}=
1.2\times 10^{-2}.
\end{eqnarray}
While the bounds on the mass insertions from saturating
\( a_{J/\psi K_{S}} \) by chargino --up squark loops for chargino mass of
order 200 GeV and light stop mass of order 200 GeV are given by \cite{emidio} 
\begin{eqnarray}
&&\sqrt{|\mathrm{Im}\left[ (\delta ^{u}_{RL})_{31}^{2}\right] |}\simeq 
4\times 10^{-1},~~~~~~~~~~~~~~~
\sqrt{|\mathrm{Im}\left[ (\delta ^{u}_{LL})_{31} (\delta ^{u}_{RL})_{31}\right]|}
\simeq 2.2\times 10^{-1},\\
&&\sqrt{|\mathrm{Im}\left[ (\delta ^{u}_{LL})_{31} (\delta ^{u}_{RL})_{32}\right]|}
\simeq 4.8\times 10^{-1},~~~~
\sqrt{|\mathrm{Im}\left[ (\delta ^{u}_{RL})_{31}(\delta ^{u}_{RL})_{32}\right]|}
\simeq 6\times 10^{-1}
\end{eqnarray}

In the framework of hierarchal Yukawa couplings we find that these
mass insertions are at least two orders of magnitude smaller than
the above limits and therefore the SUSY contributions to the CP asymmetry
\( a_{J/\psi K_{S}} \) are negligible. Thus in this scenario a real
CKM matrix is disfavoured and the only way to get large asymmetry is to have
a large \( \delta _{CKM} \) with the SM giving the leading contribution.

With democratic Yukawas however, there is a
large mixing between the different generations which can give
large mass insertions thereby accommodating the experimental
result for \( a_{J/\psi K_{S}} \) with a simultaneous saturation
of the experimental measurements of \( \varepsilon _{K} \) and \( \varepsilon' /\varepsilon  \).
Using the Yukawa textures given in eqs.(\ref{yuku},\ref{yukd}), we find that the mass 
insertion \( (\delta ^{u}_{RL})_{32} \) gives the dominant chargino contribution
to \( a_{J/\psi K_{S}} \), while the other mass insertions are at
least two order of magnitude smaller than the required bounds.\\

\begin{figure}[t]
\begin{center}
\epsfig{figure=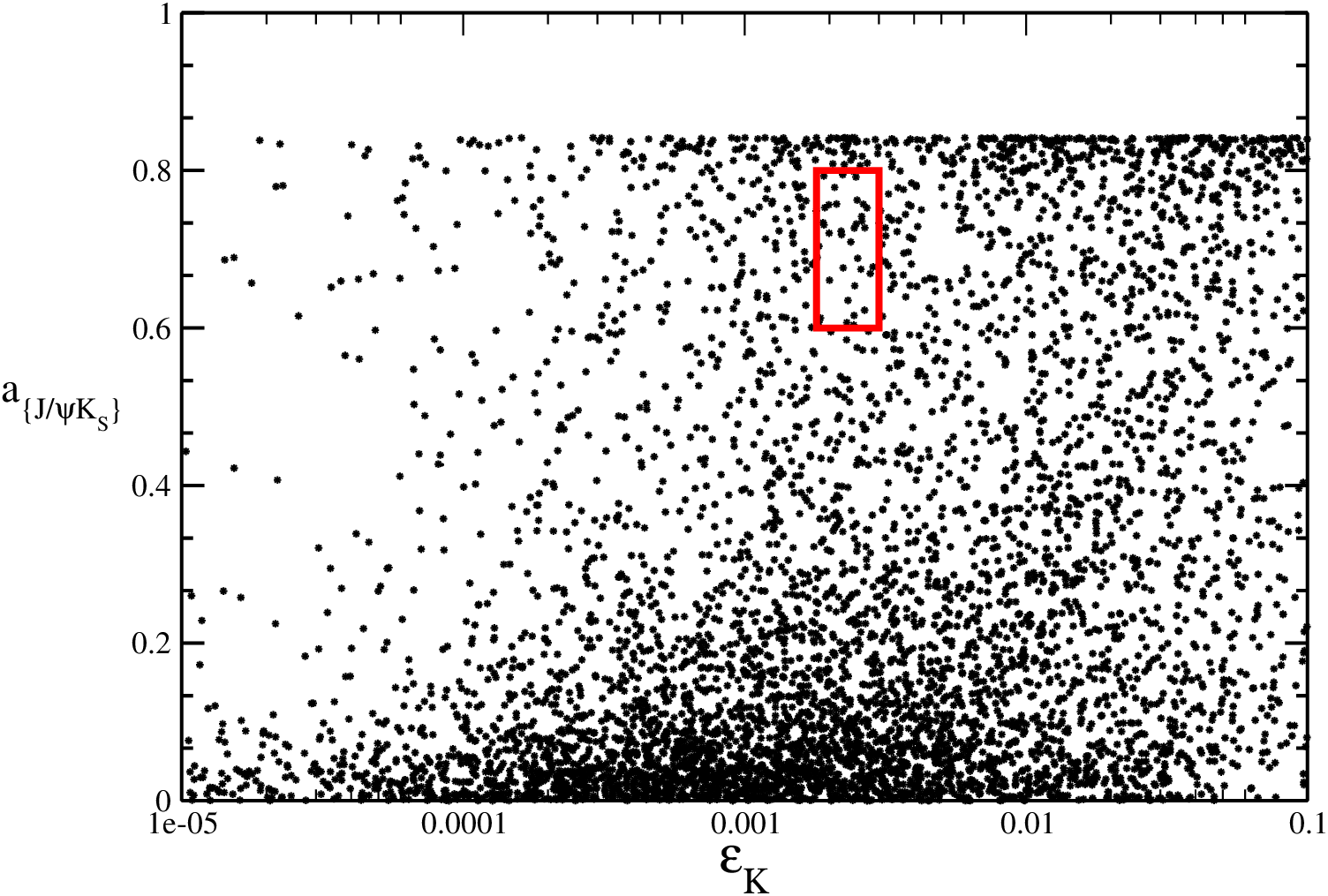,height=6.25cm,width=10.5cm,angle=0}
\end{center}
\caption{The CP asymmetry \protect\( a_{J/\psi K_{S}}\protect \) versus
$\varepsilon_K$. The SUSY parameters are fixed as in figure 1. Again, the box 
represents the SM and the constrained MSSM results.} 
\end{figure}

In figure 3 we present a scatter plot of the \( a_{J/\psi K_{S}} \)
versus the values of the parameter $\varepsilon_K$ for 
\( \tan \beta =5 \), \( m_{1/2}= 250 \) GeV and 
the off--diagonal phases \( \phi _{A_{ij}} \lsim 0.1\) in order to satisfy the 
EDM constraints. We vary the absolute values of \( A_{ij} \) from \( -3m_{0} \) 
to \( 3m_{0} \). From this figure, we see that in this class of models 
\( a_{J/\psi K_{S}}\) can be within the experimental range even with 
a real CKM matrix and small SUSY phases.

At this stage we should comment on the question of whether having a flat
unitarity triangle is consistent with present experimental data. Within the
framework of the SM, one can show that the present experimental values of
$|V_{ud} V_{ub}|$, $|V_{cd} V_{cb}|$, and $|V_{td} V_{tb}|$ are
 inconsistent with a flat unitarity triangle \cite{buras}.
However this conclusion is no longer valid in the framework of physics
beyond the SM, in particular in the supersymmetric extensions we are
considering.
This is because, while the extraction of $|V_{ub}|$ and
$|V_{cb}|$ from $B$--meason decay rates remains valid even in the
presence of New Physics (NP), the extraction of $|V_{td} V_{tb}|$ from
experimental data on $B_d - \bar{B}_d$ and $B_s - \bar{B}_s$ mixings has
to be modified \cite{GB} in order to take into account NP contributions to
the above processes.


\section{{\large \bf On approximate CP violation}}

We have seen in the previous section that supersymmetry can provide
the main contribution to CP violation with real CKM and with small
SUSY phases if the Yukawa couplings are nearly democratic and the
\( A \)--terms are non--universal.
The large flavour mixing in this model is crucial to 
compensate for the smallness of the CP phases.
In figure 4, we plot the values of $\varepsilon_K$, $\varepsilon'/\varepsilon$ 
and $\sin 2 \beta$ for $\phi_{12}=\phi_{13}=\phi_{23}=10^{-2}$, $m_0=m_{1/2}=250$ GeV 
and $A_{ij}$ vary from $-3m_0$ to $3 m_0$. As can be seen from  this figure, 
the experimental values of these quantities can be saturated by the supersymmetric 
contributions with small phases. It worth noting that due to the hermiticity 
assumption of the $A$-terms there is a severe cancellation between $LR$ and $RL$ mass 
insertions in the gluino contribution to $\varepsilon'/\varepsilon$.
Therefore, in this class of models,  the chargino contribution is the only source and hence
the values of $\varepsilon'/\varepsilon$ are smaller than usual. With 
different patterns for the $A$-terms, such as the factorizable form given 
in eq.\ref{yuku},\ref{yukd}, such 
cancellation does not occur and the gluino gives the dominant contribution resulting in
larger values for $\varepsilon'/\varepsilon$.\\

\begin{figure}[t]
{\centering \begin{tabular}{cc}
\resizebox*{0.45\textwidth}{0.2\textheight}{\includegraphics{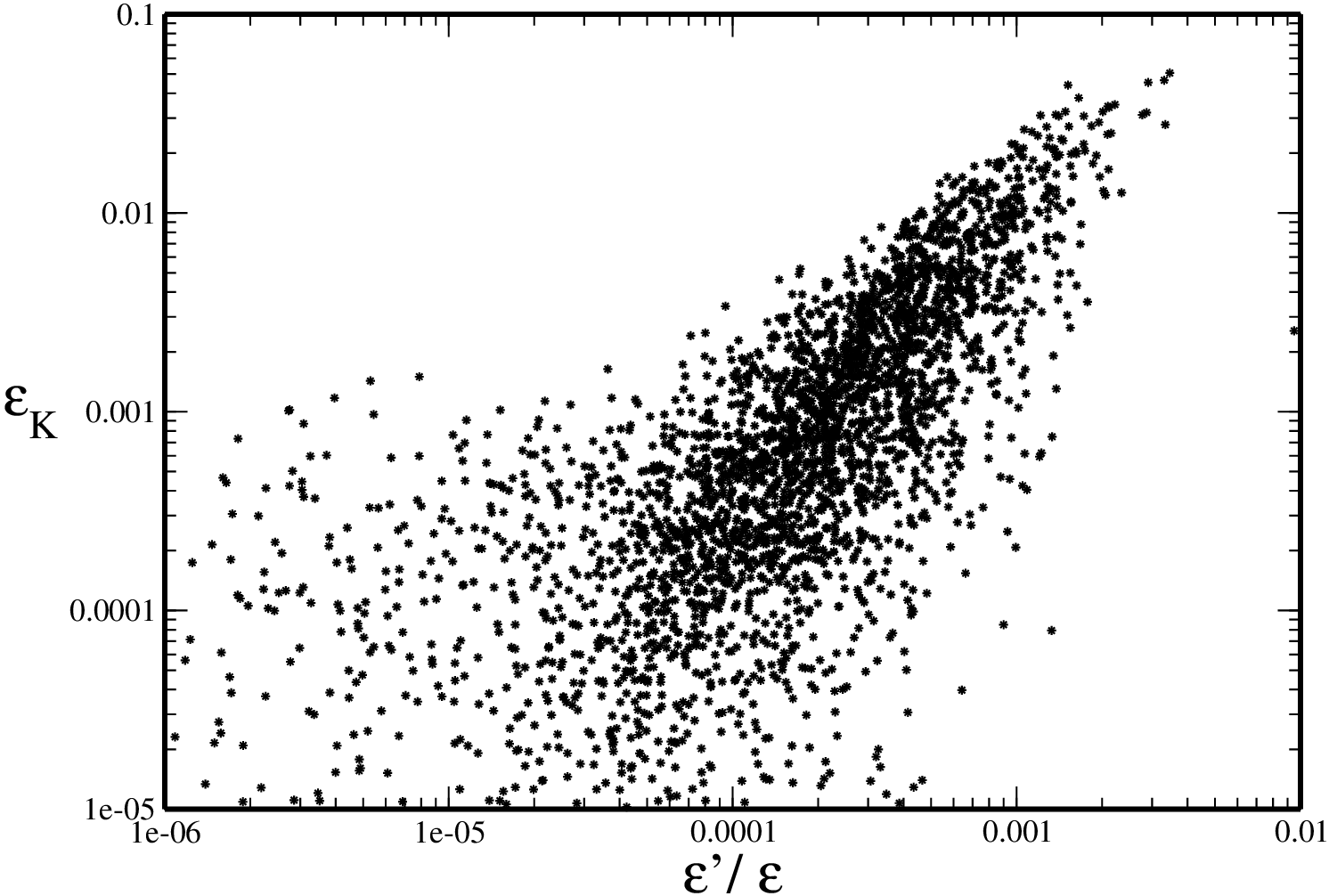}} &
\resizebox*{0.45\textwidth}{0.2\textheight}{\includegraphics{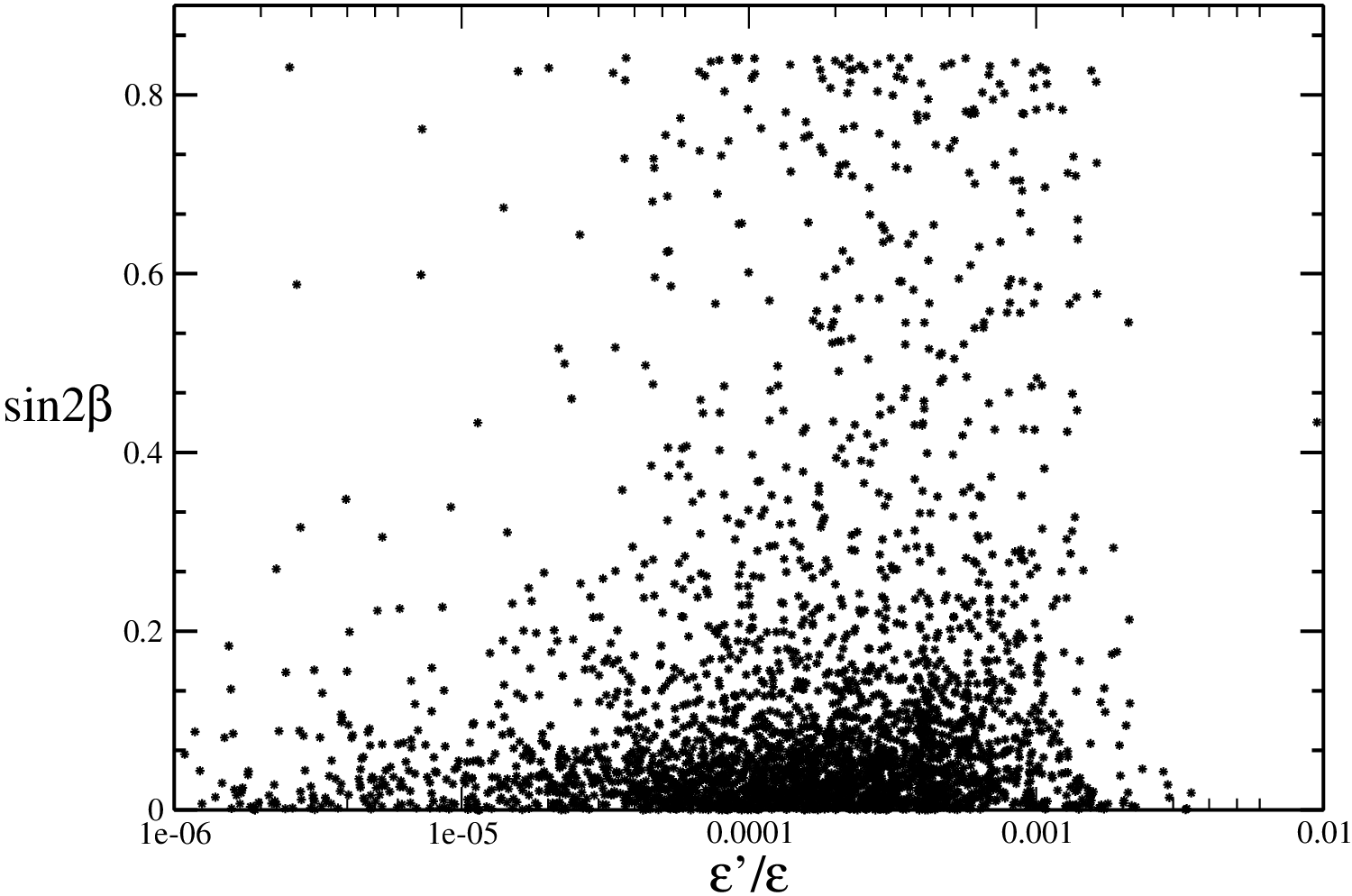}} \\
\end{tabular}\par}
\caption{The values of $\varepsilon_K$ verses $\varepsilon'/\varepsilon$ 
(left) and $\sin 2\beta$ (right) for SUSY phases equal $10^{-2}$, $m_0=m_{1/2}
=250$ GeV and $A_{ij} \in\left[-3 m_0, 3 m_0\right]$.}
\end{figure}

Since the SUSY phases are of order
$10^{-2}$ (and $\delta_{CKM}=0$), one may consider this model as an example 
of approximate CP violation. 
As discussed in section 2, approximate CP is an interesting possibility 
for solving the SUSY EDM problem. However, when the flavour structure 
is highly non-universal, it is very important to specify
at which scale the approximate CP assumption is imposed. It is clear that 
imposing it at the electroweak (EW) scale is not viable, since 
all contributions to CP violating processes would then be too 
small to accommodate the experimental 
measurements of $\varepsilon_K$, $\varepsilon'/\varepsilon$ and 
$a_{J/\psi K_S}$. However, the imposition of approximate CP only 
makes sense at the energy scale at which CP is broken (the GUT scale for example) 
then, as shown in the previous sections, all the CP violating 
phenomena at low energy scale can be accommodated thanks to renormalization group effects.  

It is worth noting in detail how the running affects the various parameters
that enter. The evolution of the CKM matrix from the GUT scale 
to the EW scale changes the value of $\delta_{CKM}$ only slightly, 
and a $\delta_{CKM}$ 
of order $10^{-2}$ at GUT scale remains of that order at the EW scale.
For SUSY parameters the situation is different however. Some quantities, 
such as the diagonal elements of $A$--terms,
receive real contributions from the gaugino masses, so that 
their phases are diluted by the running, and 
at the EW scale the phases of $A_{ii}$ are smaller than those at GUT scale. 
On the other hand, the phases of the off--diagonal $A$--terms and the 
$\mu$--term are essentially preserved by the running. 
Furthermore, with diagonal soft scalar 
masses at GUT scale we have $(M_Q^2)_{ij}=0$ for $i\neq j$. Since the running
of these quantities depends on the off diagonal elements of the $A$--terms, 
they receive some small complex contributions. However, 
the real parts of these quantities are of the same order as their
imaginary parts and hence, by the time we reach the weak scale, 
the phases of these parameters are of order one.  

In the super-CKM basis, the mass insertions $(\delta^{d,u}_{LL})_{ij}$ which
are relevant for $\varepsilon_K$ (in particular $(\delta^{d,u}_{LL})_{21}$)
and $a_{J/\psi K_S}$ (in particular $(\delta^{d,u}_{LL})_{31}$) are given by
$$ (\delta^{d,u}_{LL})_{ij} = (S^{u,d}_L M_Q^2 S_L^{u,d^{\dagger}})_{ij}.$$
It turns out that $\mathrm{arg}\left((\delta^{d,u}_{LL})_{ij}\right)\simeq
10^{-1} - 1$. However the magnitudes of these mass insertions strongly depend
on the values of the transformation matrices $S^{u,d}_{L,R}$. In the case of 
hierarchical Yukawas, these matrices have very small mixing, similar to the 
CKM matrix), and therefore the absolute value of for instance  
$(\delta^{u}_{LL})_{31}$ is of order $10^{-3}$ which is too small to account 
for the large value of $\sin 2 \beta$. On the other hand, 
with democratic Yukawas, 
the matrices $S^{u,d}$ have large mixing as in eqs.(\ref{sud}) 
which enhances the magnitude of these mass insertions, so that $\sqrt{\vert
\mathrm{Im}(\delta^u_{LL})_{31}^2 \vert}$ is of order $10^{-1}$ which is 
the required value \cite{emidio} in order to get $\sin 2 \beta \simeq 0.79$.  
The same reasoning can be given for the enhancement of 
the value of $\varepsilon_K$ with
democratic Yukawas.

We thus conclude that, due to the large SUSY CP violating phases given by
$\mathrm{arg}\left((\delta^{d,u}_{LL})_{ij}\right)$, CP is significantly 
violated at low energy scale even if it was an approximate symmetry at high
energy scale.

\section{{\large \bf Conclusion}}

To summarize, we have shown that a democratic Yukawa structure together with
flavour non-universality in the soft supersymmetry breaking, can lead to significant 
supersymmetric contributions to CP violating processes. Indeed it is still possible to
saturate all CP violating processes entirely with supersymmetry contributions even if the 
phase of the CKM matrix is small. This result contrasts sharply with the 
usual assumption that a large measured value of $\sin 2 \beta $ implies
a large CKM phase. We stress that implicit in this assumption is 
a hierarchical flavour structure and/or flavour universality. With more democratic 
flavour structures for the Yukawas and {\em generic} and non-universal SUSY breaking, we 
once again find that a small $\delta_{CKM} $ is possible.

Moreover we noted that entirely supersymmetric CP violation is still consistent 
with the notion of approximate CP ({\em i.e.} small phases). The crucial point here 
is that, when there is a large amount of flavour non-universality, renormalization groups effects 
generally turn a vertex insertion with small CP phases at a high energy scale into one with 
large phases at the weak scale. Approximate CP can therefore still be consistently imposed
at some high scale, and indeed some of the nice features of approximate CP, such as 
small $\mu$--term phases, survive the renormalization.

What does this result mean for our understanding of CP violation?  
First it shows us that it is still rather too early to dismiss alternatives to a
large CKM phase; we do not believe that a phase in the CKM matrix has yet been proven
to exist. However, from the figures presented in the paper, we clearly see that 
the standard CKM picture of CP violation is rather successful compared to 
the democratic Yukawa models presented here. Thus the second role of our analysis is 
to help quantify the success of the standard CKM model.

%


\begin{thebibliography}{99}
\bibitem{babar}
BABAR Collaboration, B. Aubert {\it et al.},
Phys.\ Rev.\ Lett. {\bf 87} 091801 (2001).\\
%
BELLE Collaboration, K. Abe {\it et al.},
Phys.\ Rev.\ Lett. {\bf 87} 091802 (2001) .
\bibitem{nir}
Y.~Nir, arXiv:hep-ph/0109090.
\bibitem{kobayashi}
S.~Khalil, T.~Kobayashi and A.~Masiero,
Phys.\ Rev.\ D {\bf 60}, 075003 (1999).
\bibitem{branco}
G.~C.~Branco, M.~E.~Gomez, S.~Khalil and A.~M.~Teixeira,
arXiv:hep-ph/0204136.
\bibitem{Abel:2001vy}
S.~Abel, S.~Khalil and O.~Lebedev,
Nucl.\ Phys.\ B {\bf 606}, 151 (2001).
\bibitem{kane}
M.~Brhlik, L.~L.~Everett, G.~L.~Kane, S.~F.~King and O.~Lebedev,
Phys.\ Rev.\ Lett.\  {\bf 84}, 3041 (2000).
\bibitem{bound} P.G. Harris {\it et al.}, Phys. Rev. Lett. {\bf 82} (1999), 904;
\bibitem{eedm} E.D. Commins {\it et al.}, Phys. Rev. {\bf A50} (1994), 2960.
\bibitem{mercury} M.V. Romalis, W.C. Griffith, and E.N. Fortson,
Phys.\ Rev.\ Lett.\ {\bf 86}, 2505 (2001);
J.P. Jacobs {\it et al.}, Phys. Rev. Lett. {\bf 71} (1993), 3782.
\bibitem{Abel:2001cv}
S.~Abel, S.~Khalil and O.~Lebedev,
arXiv:hep-ph/0112260.
\bibitem{Abel:2000hn}
S.~Abel, D.~Bailin, S.~Khalil and O.~Lebedev, Phys.\ Lett.\ B {\bf 504}, 241 (2001).

\bibitem{Khalil:2002jq}
S.~Khalil, arXiv:hep-ph/0202204.
\bibitem{Khalil:2000ci}
S.~Khalil, T.~Kobayashi and O.~Vives, Nucl.\ Phys.\ B {\bf 580}, 275 (2000)
\bibitem{Abel:1996eb}
S.~A.~Abel and J.~M.~Frere, Phys.\ Rev.\ D {\bf 55}, 1623 (1997);
S.~Khalil and T.~Kobayashi,
Phys.\ Lett.\ B {\bf 460}, 341 (1999);
K.~S.~Babu, B.~Dutta and R.~N.~Mohapatra,
Phys.\ Rev.\ D {\bf 61}, 091701 (2000);
M.~Brhlik, L.~L.~Everett, G.~L.~Kane, S.~F.~King and O.~Lebedev,
Phys.\ Rev.\ Lett.\  {\bf 84}, 3041 (2000)
\bibitem{emidio}
E.~Gabrielli and S.~Khalil, arXiv:hep-ph/0207288, to be appeared in Phys. Rev. D.
\bibitem{buras}
For a recent review, see A. Buras,  
Invited talk at 14th Rencontres de Blois, hep-ph/0210291.
\bibitem{GB}
F.~J.~Botella, G.~C.~Branco, M.~Nebot and M.~N.~Rebelo,
arXiv:hep-ph/0206133, and references therein.
\end{thebibliography}
\end{document}